\documentclass{iau} 
\usepackage{graphicx}

\title[IAUS\,352 -- NIRSpec spectroscopy on JWST] 
{Spectroscopy with the JWST Advanced Deep Extragalactic Survey (JADES) - the NIRSpec/NIRCAM GTO galaxy evolution project}

\author[Andrew Bunker]   
{Andrew J.\ Bunker$^1$
 on behalf of the NIRSpec Instrument Science Team 
  and the JADES collaboration
 }

\affiliation{$^1$Department of Physics, University of Oxford, 
Keble Road, Oxford OX13RH, United Kingdom \\ email: {\tt andy.bunker@physics.ox.ac.uk} 
}

\pubyear{2019}
\volume{352}  
\jname{Title of your IAU Symposium}
\editors{A.C. Editor, B.D. Editor \& C.E. Editor, eds.}
\begin{document}

\maketitle

\begin{abstract}
I present an overview of the JWST Advanced Deep Extragalactic Survey (JADES), a joint program of the JWST/NIRCam and NIRSpec Guaranteed Time Observations (GTO) teams involving 950 hours of observation. We will target two well-studied fields with excellent supporting data (e.g., from HST-CANDELS): GOODS-North and South, including the Ultra Deep Field. The science goal of JADES is to chart galaxy evolution at $z>2$, and potentially out to $z>10$, using the rest-frame optical and near-IR though observations from $\approx 1-5\,\mu$m. Multi-colour NIRCam imaging with 9 filters will enable photometric redshifts and the application of the Lyman break technique out to unprecedented distances. NIRSpec spectroscopy (with spectral resolving powers of $R=100$, $1000$ \& $2700$) will measure secure spectroscopic redshifts of the photometrically-selected population, as well as stellar continuum slopes in the UV rest-frame, and hence study the role of dust, stellar population age, and other effects. Measuring emission lines can constrain the dust extinction, star formation rates, metallicity, chemical abundances, ionization and excitation mechanism in high redshift galaxies. Coupling NIRCam and NIRSpec observations will determine stellar populations (age, star formation histories, abundances) of galaxies and provide the information to correct their broad-band spectral energy distribution for likely line contamination. Potentially we can search for signatures of Population III stars such as HeII. We can address the contribution of star-forming galaxies at $z>7$ to reionization by determining the faint end slope of the luminosity function and  investigating the escape fraction of ionizing photons by comparing the UV stellar continuum with the Balmer-line fluxes. 
\end{abstract}

\firstsection 
\section{Introduction}

In recent years, the high redshift frontier has greatly expanded thanks largely to upgraded cameras on HST, able to see further into the near-infrared, and follow-up spectroscopy with 8-10m class telescopes on the ground. Multi-wavelength imaging using broad-band filters enables redshifts to be estimated from the shape of the spectral energy distribution (SED), and one of the cleanest and best examples of these ``photometric redshifts" is the Lyman break technique. This relies on the absorption at Lyman-$\alpha$ (1216\AA ) produced by individual clouds of neutral hydrogen along the line-of-sight, and this Lyman-$\alpha$ forest absorption is strong at high redshift, meaning that there is a sharp drop in flux at wavelengths below Lyman-$\alpha$ from a distant galaxy or QSO, and essentially no flux below the Lyman limit at 912\AA . We can identify candidates in different redshift slices by looking at which broad-band filter a galaxy ``disappears" in due to the strong redshifted spectral break. There is a danger that objects with intrinsically red colours in the optical/near-infrared, such as low-mass stars within our own galaxy, and old or dust-reddened galaxies at intermediate redshift, can mimic the Lyman break in the UV continuum of a high redshift galaxy. However, using the colours from filters longward of the putative break can remove these low-redshift interlopers to some extent.

Going beyond simply identifying candidate high redshift galaxies, we can use the luminosities (inferred from the apparent magnitudes and estimated redshifts) and the surface densities, coupled with the depth surveyed in redshift space, to determine the luminosity function. Currently, most of our knowledge of the highest redshift galaxies is restricted to the rest-frame UV (just above the 1216\AA\  break). The UV luminosity is usually dominated by the hot, massive, short-lived OB stars, and hence is a proxy for the star formation rate, although this is subject to the stellar initial mass function and to dust obscuration. 

Unfortunately, most of the UV luminosity functions at high redshift are based just on candidate Lyman Break Galaxies (LBGs) from broad-band imaging alone, with no spectroscopic confirmation. Accurate spectroscopic redshifts are critical, since the interloper fraction of lower-redshift sources is very uncertain, and also the inferred luminosity function depends strongly on the calculation of the volume sampled, which in turn depends on the selection and completeness which are strongly affected by the redshift distribution and spectral slopes of the LBGs. Current UV luminosity functions are uncertain, but for now obtaining spectroscopic redshifts at $z>6$ is supremely challenging, since current spectroscopy is largely limited to the rest-frame UV, and the most notable emission line, Lyman-$\alpha$, is often weak or absent at high redshift, perhaps through scattering of this resonant line by neutral gas in the IGM. While a minority of the $z\sim 6$ candidates have had secure spectroscopic redshifts from Lyman-$\alpha$ (e.g., Bunker et al. 2003, Stark et al. 2010), beyond $z\approx 7.5$ the success in confirming the photometric redshifts through
spectroscopy is very low (e.g. Caruana et al. 2014), with some rare exceptions (e.g. Zitrin et al. 2015 at $z=8.68$). Spectral breaks have been seen in low-dispersion slitless spectroscopy (e.g., Oesch et al.\ 2016 GN-z11 which may be at $z\approx 11$), confirming the broad-band photometry but not giving a definitive redshift from emission or absorption lines. While some progress has been made using other weak rest-UV lines such as CIII]1909 (Stark et al. 2015) and also in the far-IR with ALMA (e.g., [CII] $158\,\mu$m, Smit et al. 2018), the next critical step is spectroscopy of the rest-frame optical where there are many well-understood emission lines. This will be accomplished by the near-infrared spectrograph, NIRSpec, on the James Webb Space Telescope, scheduled for launch in 2021.

\section{NIRSpec -- the Near-Infrared Spectrograph on JWST}

NIRSpec operates in the range $0.6-5\,\mu$m, and has three spectral resolutions: a low-dispersion prism ($R=100$) which captures all the wavelength range with a single exposure, and medium- and high-resolution gratings ($R=1000$ and $R=2700$) which use 3 bands to cover the wavelength range. The unique feature of this spectrograph is its use of micro-shutter arrays, developed specifically for NIRSpec to enable multi-object spectroscopy. Each $100\times 200\,\mu$m micro-shutter subtends $0.2''\times 0.4''$ and can be individually commanded to open, and they are arranged in 4 arrays each containing $171\times 365$ micro-shutters. By opening shutters on targets, we can significantly reduce the background intensity (from zodiacal light and from other astronomical sources in the field), essentially building a slitmask in space and becoming far more sensitive than slitless spectroscopy. The NIRSpec field of view covers over $3'\times 3'$ with an unvignetted area of 5.5 square arcminutes, well matched to existing deep fields and the JWST NIRCam imager. There is also a $3''\times 3''$ integral field unit with $0.1''$ sampling for spatially-resolved spectroscopy of individual objects, along with a number of fixed slits for traditional long-slit work.

\section{The JWST Advanced Deep Extragalactic Survey (JADES) GTO Observations}

The NIRSpec Instrument Science Team (IST) will undertake a cohesive ``wedding cake" survey in our GTO time, targetting $z>7$ galaxies in our deepest fields (potentially out to $z\sim 20-30$) and $1<z<7$ galaxies in our Medium and Wide fields. We intend to make a genuine impact on our understanding of galaxy evolution, rather than spreading the GTO time over a number of disparate projects. Since July 2015 we have been collaborating with the NIRCam Instrument Science Team (the PI for which is Marcia Rieke at the University of Arizona -- see her contribution to the proceedings of this IAU Symposium), and the NIRCam and NIRSpec ISTs have an integrated extra-galactic programme (JADES).

In collaboration with our NIRSpec-IST, the NIRCam-IST will spend 450 hours imaging our NIRSpec survey fields with 7 broad-band filters spanning $0.9-4.4\,\mu$m, to $10\,\sigma$ point-source depths of $AB=29.5-29.8$\,mag in the Deep fields, and $AB=28.6-29.0$\,mag in the Medium fields. We will use these images, along with the existing deep HST images at shorter wavelength (as well as other data such as from ALMA), to select targets for NIRSpec-IST spectroscopy. The Deep Tier has been allocated 150 hours of GTO time. We cover two pointings within the GOODS-South field, one of which contain the Hubble Ultra Deep Field (HUDF) region, also known as the eXtreme Deep Field (XDF). We will obtain 25\,ksec using each of the three medium-dispersion ($R=1000$) gratings, spanning $1-5\,\mu$m, and the high resolution F290LP/G395H mode. The spectral resolution is sufficient to resolve the [NII]6583 doublet from H$\alpha$ for emission line diagnostics, and to measure line widths (for example the asymmetry of Lyman-$\alpha$ emission due to blue-wing absorption). Our simulations indicate that we can allocate 60--100 targets on the micro-shutter assembly (MSA), depending on target density and allowing three microshutters for each object (dithering up and down to improve background subtraction). We will also use the low-dispersion $R=100$ prism for 100\,ksec at each pointing, which captures the full wavelength range $0.8-5\,\mu$m and has greater sensitivity to continuum emission than $R=1000$, and because the prism spectra are short compared to the width of the two NIRSpec 2K infrared arrays we can potentially have 3--4 targets on each spectral row, increasing our multiplex to $\sim$200--300 targets per pointing. All these observing modes have 10$\sigma$ sensitivities to line emission of $5-9\times 10^{-19}\,{\rm erg\,cm^{-2}\,s^{-1}}$ at $\lambda>2\,\mu$m.

The Medium Tier of the survey will use the same spectral set-ups, but with integration times 20\% those of the Deep (i.e., sensitivities of about half the Deep Tier where the observations are zodiacal background limited). The Medium Tier has been allocated 200 hours of GTO time to target 12 pointings distributed over both the GOODS-North and GOODS-South fields, and we supplement this with a Wide survey over some of the rest of the GOODS/CANDELS fields, taking snapshots of 3\,ksec with the R=100 prism, and 2\,ksec each with the two redder $R=2700$ gratings ($2-5\,\mu$m). Having three tiers of increasing area and decreasing sensitivity means that we can sample a range in the luminosity function with comparable numbers of galaxies in each of the tiers (three luminosity bins). The Wide survey will do 35 pointings for a total of 100 hours, and will focus on galaxies at intermediate redshift ($1<z<5$) and rare bright targets at $z>5$. The NIRSpec-IST three-tier survey with the MSA has a total of 470 hours of GTO time, and we will spend a further 270 hours using the Integral Field Unit with the high-dispersion $R=2700$ gratings to obtain spatially-resolved spectra of $\approx$50 high redshift objects, to derive kinematics (rotation curves and outflows) and metallicity gradients. Our final high redshift targets are the most distant known QSOs at $z>6.7$, each of which we will target for 5 hours on source with the NIRSpec fixed slits and high resolution grating to determine the IGM opacity and study the Lyman-$\alpha$ damping wing, as a probe of reionization.

\section{Emission lines from High Redshift Galaxies}

NIRSpec will deliver multiple emission line measurements for high redshift galaxies in the rest-frame optical and UV (Ly$\alpha$, H$\alpha$+[NII], [SII], H$\beta$, [OII]3727\AA , [OIII]5007\AA\ etc.), enabling us to use ``BPT" diagrams (Baldwin, Philips \& Terlevich 1981) to address the nature of the photoionization in individual galaxies (i.e. star formation vs. AGN). Diagnostic ratios will also give the gas-phase metallicity, and we can apply the popular ``R23" measure of O/H (Pagel et al. 1979, using [OII], [OIII] and H$\beta$) out to $z\approx 10$ with NIRSpec (Chevallard et al. 2019). At $z<7$ with the $R=1000$ grating, we can use the [NII] line to break the ``double fork" degeneracy in the plot of R23 against metallicity. From a wider range of line ratio diagnostics we can potentially get individual abundances for C, N and O, and these abundance patterns should evolve differently with redshift due to the different timescales involved in the production of the elements. 

One of the most ambitious science goals of JWST is the discovery of the first generation of stars, forming from the pristine intergalactic medium of hydrogen and helium from Big Bang nucleosynthesis, before contamination by heavier elements produced in stellar nucleosynthesis. Some simulations predict very massive stars due to a lack of metal cooling lines, with a hard spectrum that can doubly-ionize helium. An expected signature of Population III is the HeII1640 emission line, and the absence of metal lines.  The search for Population III is challenging, but we can improve our chances of detecting HeII (and tighten the limits on metal lines) by stacking our JWST spectra of faint galaxies at Lyman-$\alpha$.

\section{Exploring Reionization with JWST}

Recent results indicate the mid-point of reionization may have occurred at $z\approx 8-9$ (Planck collaboration, 2016). The source of necessary ionizing photons remains an open question: the number density of high redshift quasars is insufficient at $z > 6$ to achieve this 
(Dijkstra et al. 2004). There has been speculation that low-luminosity AGN might contribute more ionizing photons (Giallongo et al. 2015). If these are indeed numerous enough to account for reionization then they will be identified as NIRSpec targets in our Lyman break selection, and the line ratio diagnostics will reveal their AGN nature. Star-forming galaxies at high redshift are a more likely driver of reionization, but we must first determine their rest-frame UV luminosity density. Other important and poorly-constrained factors are the slope of their UV spectra (and hence the number of ionizing photons) and the escape fraction of ionizing photons from these galaxies ($f_{esc}$, the fraction of ionizing photons formed on the photospheres of OB stars which reach the low-density IGM).

The star-forming galaxies we see at $z\approx 6$, even to the limiting depth of the HUDF ($\approx 28.5$\,mag in $z$-band, corresponding to an absolute magnitudes $M_{UV}\approx -17$), do not provide enough photons to maintain the ionization of the Universe, even if the escape fraction is 100\% (Bunker et al. 2004). The short-fall is even more apparent around the reionization epoch at $z\approx 8$ (Lorenzoni et al. 2013). Undoubtedly galaxies fainter than the detection limit contribute, and current indications are that the luminosity function may have a steep faint-end slope at high redshift (perhaps a Schechter function with $\alpha\simeq -2$), but even so we need to extrapolate well beyond current observational limits to fainter luminosities for there to be sufficient photons.

Our JWST GTO programme will go $1-2$\,mag.\ fainter in the UV luminosity function with LBGs from NIRCam imaging, as well as using NIRSpec spectra for accurate redshifts and better template SEDs to improve the accuracy of the UV luminosity functions at $z>6$, which are highly uncertain at high redshift. Another huge uncertainty is the escape fraction of ionizing photons, and while some  measurements exist at $z\sim 3$ 
the high optical depth of the IGM at $z>6.3$ means that we may never directly observe the ionizing photons at these high redshifts. We can use the indirect method of Zackrisson et al. (2013) on our NIRSpec spectra to estimate the escape fraction of ionizing photons by comparing H$\alpha$ emission with the extrapolated UV continuum shortward of 912\AA\  (corrected for dust by the Balmer decrement) out to $z\sim 7$, and using H$\beta$ out to $z\sim 10$. The H$\alpha$ luminosity is tied to the number of ionizing photons which do not escape the galaxy (and are absorbed by H{\scriptsize ~I}). Extrapolating the UV continuum from the observed flux above 1216\AA\  introduces some uncertainty -- in many cases we can measure the UV spectral slope, $\beta$, from the $R=100$ low-dispersion spectrum or the broad-band photometry. Indeed the blue slopes we observe at $z>6$ in the rest-UV (Stanway, McMahon \& Bunker 2005; Wilkins et al. 2011) could be explained through low metallicity, or a top-heavy initial mass function (IMF), which can produce between 3 and 10 times as many ionizing photons for the same 1500\AA\  UV luminosity as a Salpeter IMF (Schaerer 2003).
Binary stars can further increase this correction factor (Eldridge \& Stanway 2009). However, for reionization, the important quantity is the number of escaping ionizing photons -- the product of the escape fraction of ionizing photons and the emissivity of ionizing photons per UV luminosity density longward of Lyman-$\alpha$ (usually measured around 1500\AA ), and this product is more robust than the individual quantities $f_{esc}$ and the number of ionizing photons. The ultimate goal is to measure the escape fraction (and number of escaping photons) as a function of stellar mass, star formation rate and metallicity, and also kinematics and outflows (determined by ISM line offsets from the systemic redshift). We can then address the nature of the galaxies which reionize the Universe.


\begin{thebibliography}{}
\bibitem[Baldwin et al. (1981)]{BPT1981}
Baldwin, J. A., Phillips, M. M. \& Terlevich, R. 1981 \textit{PASP}, 93, 5

\bibitem[Bunker et al. (2003)]{Bunker_etal03}
Bunker, A. J., Stanway, E. R., Ellis, R. S., et al. 2003, \textit{MNRAS}, 342L, 47

\bibitem[Bunker et al. (2004)]{Bunker_etal04}
Bunker, A. J., Stanway, E. R., Ellis, R. S., et al. 2004, \textit{MNRAS}, 355, 374 

\bibitem[Caruana et al. (2014)]{Caruana_etal14}
Caruana, J., Bunker, A. J., Wilkins, S. M., et al. 2014, \textit{MNRAS}, 443, 2831 

\bibitem[Chevallard et al. (2019)]{Chevallard_etal19}
Chevallard, J., Curtis-Lake, E., Charlot, S. et al. 2019, \textit{MNRAS}, 483, 2621

\bibitem[Dijkstra et al. (2004)]{Dijkstra_etal04}
Dijkstra, M., Haiman, Z. \& Loeb, A. 2004, \textit{ApJ}, 601, 666

\bibitem[Eldridge \& Stanway (2009)]{EldridgeStanway_etal09}
Eldridge, J. J. \& Stanway, E. R. 2009, \textit{MNRAS}, 400, 1019

\bibitem[Giallogno et al. (2015)]{Giallongo_etal15}
Giallongo, E., Grazian, A., Fiore, F. et al. 2015, \textit{A\&A}, 578, 83

\bibitem[Lorenzoni et al. (2013)]{Lorenzoni_etal13}
Lorenzoni, S., Bunker, A. J., Wilkins, S. M., et al. 2013, \textit{MNRAS}, 429, 150 

\bibitem[Oesch et al. (2016)]{Oesch_etal16}
Oesch, P. A., Brammer, G., van Dokkum, P. G., et al. 2016, \textit{ApJ}, 819, 129

\bibitem[Pagel et al. (1979)]{Pagel_etal79}
Pagel, B., 
Edmunds, M. G., Blackwell, D. E., Chun, M. S. \&
Smith, G. 1979, \textit{MNRAS} 189, 95

\bibitem[Planck Collaboration (2016)]{Planck_16}
Planck Collaboration 2016, \textit{A\&A}, 596A, 108

\bibitem[Schaerer (2003)]{Schaerer_03}
Schaerer, D., 2003, \textit{A\&A}, 397, 527

\bibitem[Smit et al. (2018)]{Smit_etal18}
Smit, R., Bouwens, R. J., Carniani, S. et al. 2018, \textit{Nature}, 553, 178

\bibitem[Stanway et al. (2005)]{Stanway_etal05}
Stanway, E. R., McMahon, R. G. \& Bunker, A. J. 2005, \textit{MNRAS}, 359, 1184 

\bibitem[Stark et al. (2015)]{Stark_etal15}
Stark, D. P., et al. 2015, \textit{MNRAS}, 450, 1846

\bibitem[Stark et al. (2010)]{Stark_etal10}
Stark, D. P., Ellis, R. S., Chiu, K. et al. 2010, \textit{MNRAS}, 408, 1628 

\bibitem[Wilkins et al. (2011)]{Wilkins_etal11}
Wilkins, S. M., et al. 2011, \textit{MNRAS}, 417, 717

\bibitem[Zackrisson et al. (2013)]{Zackrisson_etal13}
Zackrisson, E., Inoue, A. K.. \& Jensen, H. 2013, \textit{ApJ}, 777, 39

\bibitem[Zitrin et al. (2015)]{Zitrin_etal15}
Zitrin, A., Labb\'e, I., Belli, S., et al. 2015, \textit{ApJ}, 810L, 12
 

\end{thebibliography}
\end{document}